\documentclass[aps,showpacs]{revtex4}
\usepackage{color,graphicx}
\usepackage{epsfig,amsmath}

\begin{document}
\title{Neutron-rich nuclei and the equation of state of stellar matter}

\author{F. Gulminelli$^{1,2}$ }
 
\affiliation{$^{1}$~CNRS, UMR6534, LPC ,F-14050 Caen c\'edex, France\\
$^{2}$~ENSICAEN, UMR6534, LPC ,F-14050 Caen c\'edex, France\\
}

\begin{abstract}
In this contribution we will review our present understanding of the matter equation of state in the density and temperature conditions where it can be described by nucleonic degrees of freedom. At zero temperature, all the information is contained in the nuclear energy functional in  its isoscalar and isovector channels. At finite temperature, particular emphasis will be given to the specificity of the thermodynamics in the nucleonic regime, with the simultaneous presence of long range electromagnetic and short range nuclear interactions. The astrophysical  implications of the resulting phase diagram,  as well as of different observables of exotic nuclei on the neutron-rich side,  will be touched upon.
\end{abstract}

\today

\maketitle

\section{Introduction}

Atoms and molecules are the constituents of all forms of matter constituting our everyday experience. However, because of the atoms internal structure, if it is possible to conceive situations where matter would be compressed under extreme pressure of the order of $P\approx 1 MeV/fm^{3}\approx 10^{30} N/m^3$ , it is clear that matter would be governed by nucleonic degrees of freedom. Such extreme conditions are indeed produced in nature in the most violent astrophysical phenomenon we know, namely the supernova explosion induced by core-collapse in very massive stars. 

Triggered by the first data taking of X and gamma satellites, as well as by the improved capabilities of radio telescopes and high and ultra-high gamma ray detectors, there has been in the last ten years an impressive accumulation of observational data on supernova and neutron stars in different environnements and at different evolution stages.  Among the most important recent results one should mention the precise measurement  of a very massive two solar-mass neutron star\cite{demorest} which challenges a number of theoretical equations of state of dense matter, and the measurement of the cooling curve of the young pulsar Cassiopea A\cite{cassiopea}, which gives strong constraints on the superfluid properties of neutron star cores.

Undestanding these data requires a good modelization of the supernova dynamics and of the formation process of neutron stars\cite{langanke}, but also a precise and detailed control of the relevant microphysics. 
Sophisticated two and three-dimensional core-collapse supernova simulations\cite{nakazato,marek} show that matter in the supernova core explores an extremely large interval of baryonic densities, ranging from  about $\rho>10^{10}$ g $\cdot$ cm$^{-3}\approx 10^{-4}\rho_0$  
to several times the normal nuclear density $\rho_0$ , temperatures between some hundreds of KeV and around 20 MeV, 
and proton fractions between $\approx$ 0.5 and $ \approx$ 0.3. 
An exemple of the baryonic density, temperature and proton fraction distribution in the core of a 15 solar-mass progenitor star during the first 25 ms after  bounce is reported in Figure 1.
Densities in the same range and proton fractions as low as $\approx$ 0.1-0.2 are believed to be reached in the residual neutron star which is left over after the explosion.

\begin{figure}
\includegraphics[width=4.in,height=4.in,clip]{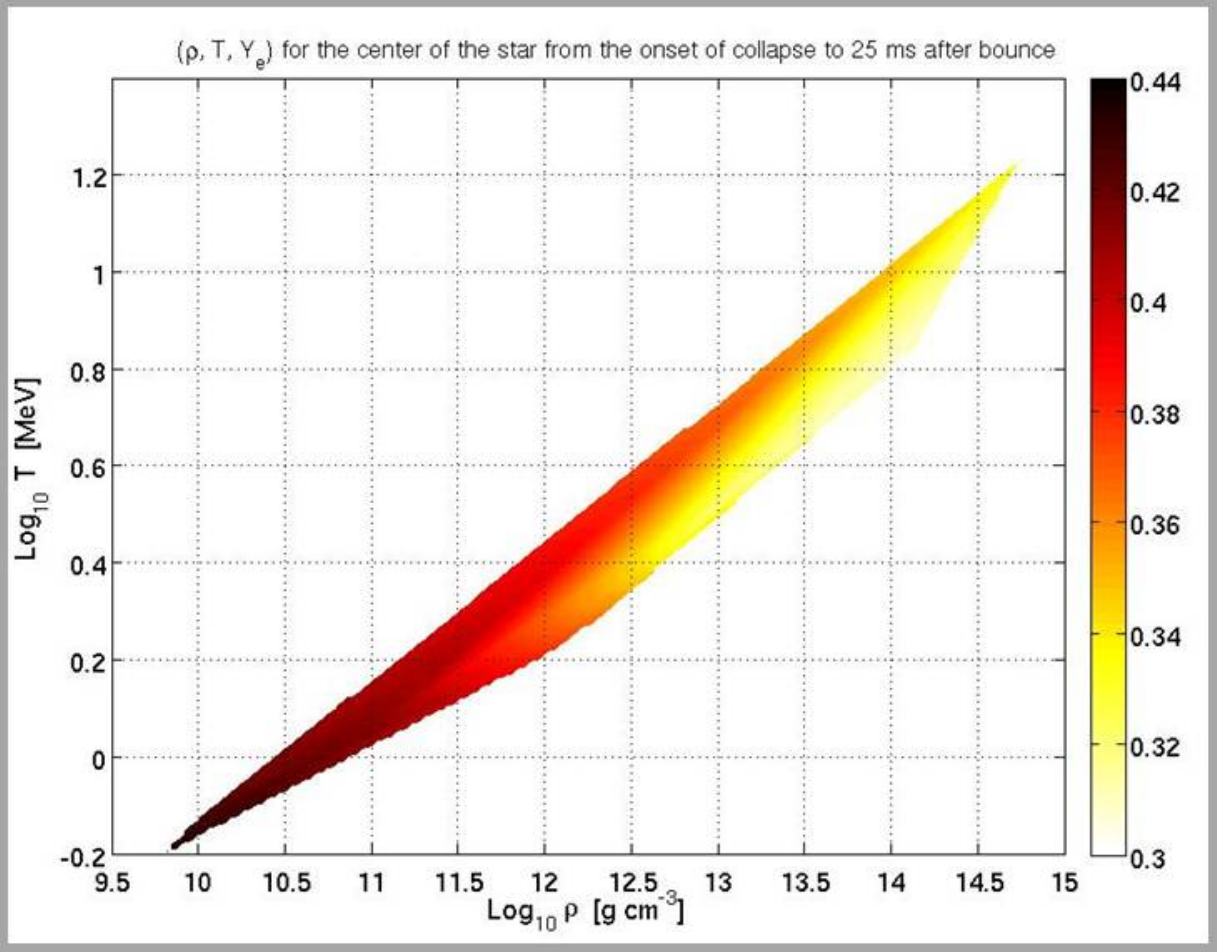}
\caption{  (Color online) Simulation of the baryonic density, temperature and proton fraction distribution in the core of a 15 solar-mass progenitor star during the first 25 ms after  bounce. Figure taken from ref.\cite{anthea}
 .}
\label{fig1}
\end{figure}

 Different microscopic properties of nuclear matter in these wide thermodynamic conditions 
are of influence for these astrophysical phenomena.
In particular, the core-collapse gravitational models show an important influence of the explosion mechanism with respect to the equation of state.  To give an example, the explosion of very massive 15 solar-mass progenitors can presently only be achieved with a so-called "soft" equation of state. As we will argue in this contribution, in the whole sub-saturation region matter is not uniform but it is constituted of finite nuclei, with a dominance of exotic neutron-rich isotopes. Therefore a reliable calculation of the abundancies of these nuclei is essential, as well as a knowledge of their mass, level densities and in-medium self-energy  modifications.  
 
Another key aspect concerns the thermal energy evacuation during the explosion and the proto-neutron star cooling. In the first evolution stage, heat is evacuated essentially by electron capture processes followed by neutrino emission. It is therefore very important to control the associated electro-weak processes, namely the electron capture rates and the neutrino interactions with the dense matter and the different nuclei of the proto-neutron star crust. In the later cooling phase the leading mechanism is conduction and the key ingredient is given by the heat capacity of the neutron star, which in turn is closely linked to the superfluid properties of matter and exotic nuclei. 
All of these microphysics properties need to be either directly experimentally measured or to be constrained by experimental measurements. 

This statement is trivial as long as masses, level densities, pairing gaps and reaction rates are concerned. Concerning quantities which are specific to the bulk limit of infinite nuclear matter, as equations of state, chemical compositions and phase structure, they have obviously to be evaluated from a theoretical model. However, as we will discuss in this contribution, these quantities cannot be calculated in a completely ab-initio theory, but depend  on models which in turn contain phenomenological parameters which need to be experimentally constrained. Relevant experimental information comprise collective modes and neutron skins in neutron-rich nuclei\cite{elias}, and transport observables in heavy-ion collisions with neutron rich nuclei. In this respect, the developement of the next-generation exotic beam facilities (SPIRAL2, FAIR and EURISOL on the long range) and the exploitation of the existing ones (RIKEN) will give essential information for our understanding of the structure and interaction of dense matter.

\section{Equation of state at zero temperature}

Nuclear matter is theoretically defined as an idealized bulk medium of neutrons and protons where the electromagnetic interaction is artificially switched off in order to achieve a thermodynamic limit.
For this to represent a realistic description of dense baryonic matter composing compact stars, different conditions have to be verified.  First, the density must be low enough for hyperonic degrees of freedom to be neglected, and moreover the system should be homogeneous, such that the net zero charge characterizing stellar matter corresponds to a net zero electromagnetic interaction. The first condition is realistic for densities below about 0.3 fm$^{-3}$, while
we will discuss the validity of the second approximation in the next section in great detail.

The functional dependence of the nuclear matter energy density on the baryonic density at a fixed value of the proton fraction is traditionally referred to as the nuclear matter equation of state.  
The first studies of the equation of state concerned only symmetric nuclear matter with equal proportion of protons and neutrons. Though such a system does not correspond to any physical phenomenon explored by nature, an empirical constraint on this functional is given by the so-called saturation point. This is defined as  a minimum in the functional located at a density corresponding to the central density of heavy nuclei, as determined by electron scattering $\rho_0 = 0.166 \pm 0.018$ fm$^{−3}$, and an energy per particle given by the bulk term of the mass formula, fitted to the available nuclear masses $ E/A = −16 \pm 1$ MeV.  
The first microscopic studies of nuclear matter date of the late seventies\cite{day,pandha} and were done,  following the pioneering work of K.A.Brueckner\cite{brueckner},  within the Brueckner-Hartree-Fock (BHF) theory or within variational approaches starting from bare two -body nucleon-nucleon interactions fitted to empirical phase shifts and deuteron data. Relativistic Dirac-Brueckner approaches (DBHF) were developed soon after\cite{malfliet}.
The present status of such models is presented in Figure 2\cite{li}.

\begin{figure}
\includegraphics[width=4.in,height=4.in,clip]{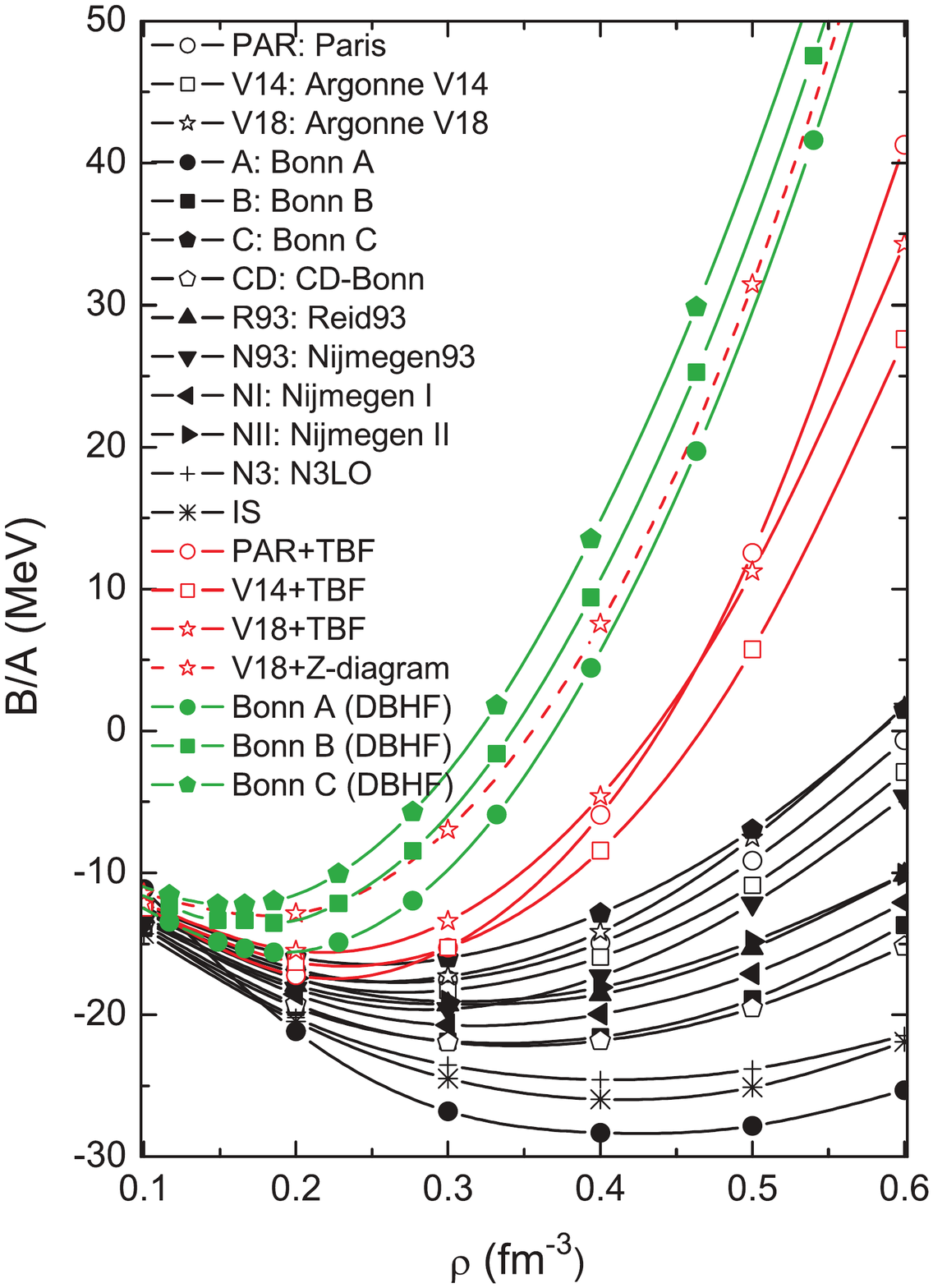}
\caption{  (Color online) Energy per baryon as a function of the matter density in recent relativistic and non-relativistic microscopic G-matrix calculations with different two and three-body nuclear interaction. Black curves: non-relativistic with two-body only. Red curves: non-relativistic with two and three-body. Green curves: relativistic calculations.
Figure taken from ref.\cite{li}. Copyright (2006) by the American Physical Society. 
 .}
\label{fig2}
\end{figure}
We can see that the different calculations vary widely, but the constraint given by the empirical saturation point bears important information on the nuclear energy functional.
Specifically, only including three-body force the BHF calculations succeed in closely reproduce the saturation point, yielding slightly less repulsive results than the DBHF results.  The relation between the relativistic and non-relativistic calculations can be understood from the fact that that the major effect of the
DBHF approach amounts to including the three-body forces corresponding to nucleon-antinucleon excitation by 2$\sigma$  exchange within the BHFcalculation. This is illustrated for the case of the V18 potential (open stars) by the dashed (red) curve in the figure, which includes only the 2$\sigma$-exchange “Z-diagram”
three-body contribution. The remaining three-body components are overall attractive and produce the final solid (red) curve in the figure. 

From this figure it is also clear that far from the saturation point the behavior of the energy functional is strongly model dependent. The overall variation of the calculations can be approximately measured by the value of the second derivative of the functional at the saturation point, the so called incompressibility coefficient $K_\infty$. In this respect, the different equations of state can be classified among extreme behaviors defined as as "soft" (less repulsive than BHF, that is with   $K_\infty\approx 200 MeV$) or "stiff" (more repulsive than DBHF, that is with  $K_\infty\approx 400 MeV$) .

An alternative phenomenological approach to the energy functional of symmetric nuclear matter 
was also proposed since the very early days of the research on the equation of state\cite{vautherin,brack,walecka}. These models are based on effective density-dependent nuclear forces or
effective interaction Lagrangians with couplings  adjusted
to fit ground state masses and charge radii over a large region of the mass table. During the years such approaches have reached a very high level of sophistication, and the predictive power of the associated effective energy functionals is extended also to the excited state properties with a level of accuracy which is not reached by any other approach in the domain of heavy and medium-heavy nuclei\cite{bender,meng}. 
 
Because of their phenomenological character, these models reproduce by construction the saturation point of symmetric nuclear matter while the incompressibility can be almost freely varied from the soft to the stiff limit.

Because of this irreducible model dependence, a major challenge for the field  has been the determination of this quantity from the comparison of experimental observables sensitive to the compression of matter to transport or structure calculations where the same functional as for the nuclear matter calculation is implemented. 

\begin{figure}
\includegraphics[width=0.48\columnwidth,height=3.5in,clip]{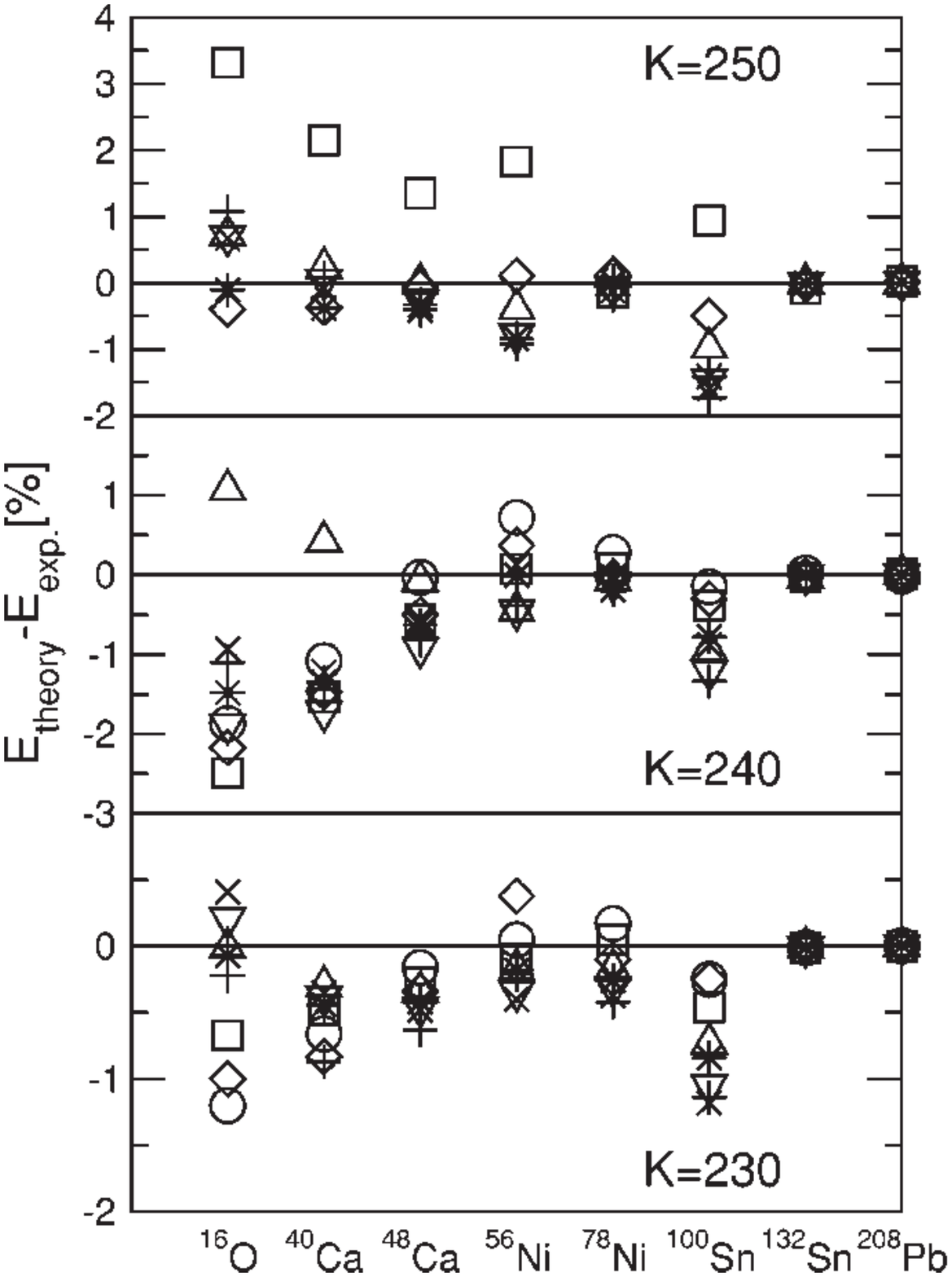}
\includegraphics[width=0.48\columnwidth,height=4.5in,clip]{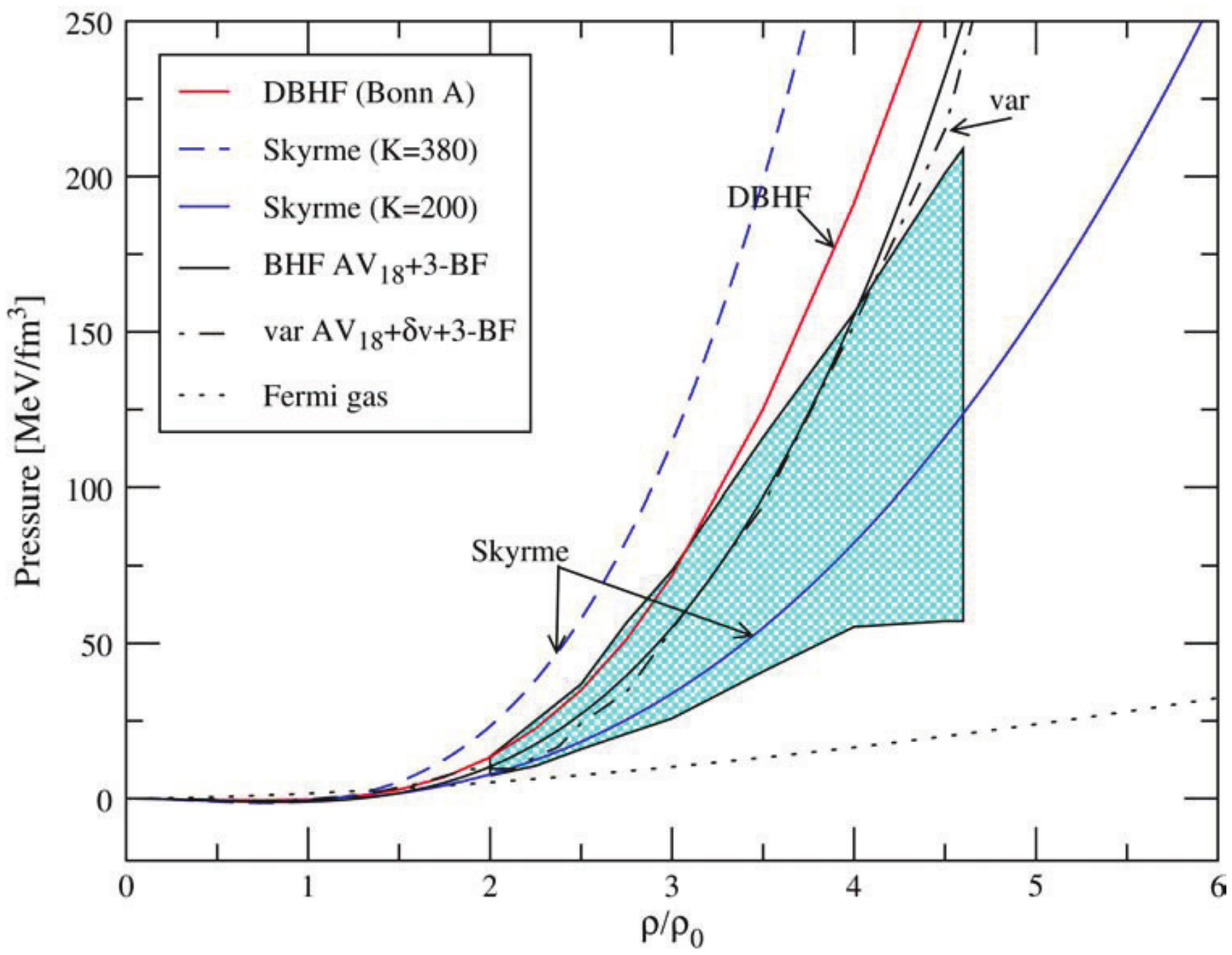}
\caption{ (Color online) Left side: difference between the experimental and calculated isoscalar giant monopole resonance centroid energies in different spherical nuclei, obtained with fully self-consistent non-relativistic mean-field calculations with Skyrme functional characterized by different incompressibility coefficients and different values for the symmetry energy at saturation.   Figure taken from ref.\cite{colo}. Copyright (2004) by the American Physical Society.  Right side: equation of state constraint from the reproduction of Au+Au transverse flow by transport calculations. Figure taken from ref.\cite{fuchs}.
 .}
\label{fig3}
\end{figure}

Through this effort for over three decades, the present incompressibility constraint can be given as $K_\infty=240\pm 30$ MeV.  This number comes from completely independent calculations of very different experimental observables. On one side, accurate fits of the monopole response of spherical nuclei of very different size have been performed with different many-body techniques and different energy functionals. 
As an example, the left part of Fig.3 shows the difference between the experimental and calculated isoscalar giant monopole resonance centroid energies in different spherical nuclei, obtained with fully self-consistent non-relativistic mean-field calculations with Skyrme functional characterized by different incompressibility coefficients\cite{colo}. Other terms of the energy functional influence the monopole response, in particular the value of the symmetry energy at saturation, which is taken as a free parameter varying between 26 (circles) and 40 MeV (crosses) in the calculations of Figure 3. The interference between the different parameters is at the origin of the error bar in the estimation of $K_\infty$.

On the other side, important independent constraints have been obtained from the comparison of heavy ion collision data at 400 MeV per nucleon to transport calculations where different energy functionals can be implemented\cite{lynch}. 
This is shown in the right part of Figure 3, which shows the density and pressure constraint obtained by such calculations in ref.\cite{lynch}. Compatible results have also been obtained 
 from subthreshold kaon production \cite{fuchs} in relativistic
nucleus-nucleus collisions.  

\section{Asymmetric matter and symmetry energy}

In the last decade, we have witnessed an impressive evolution of the microscopic and ab-initio modelizations of nuclear matter.
In particular, effective field theory, either based on density functional theory or from  chiral perturbation, has proved to be a very effective method  to provide systematic expansions of relevant diagrams which effectively separate long and short range components\cite{furnstal,weise,schwenk}.
However, the renormalization technique induces many-body forces that
have been carefully included in light nuclei but not systematically in nuclear matter, making calculations highly uncertain at high density. Alternatively, bare nucleon-nucleon forces  adjusted to reproduce the
two-body scattering and properties of light nuclei with very
high precision are used in non-perturbative calculations where  the strong correlations are
accounted  using quantum Monte Carlo\cite{gandolfi} methods. This approach is however also limited to relatively low density.

As we have discussed in the previous section, a phenomenological constraint on the density dependence is given by the incompressibility of symmetric nuclear matter. However, , since laboratory nuclei probe only
nearly isospin-symmetric matter,  the variation of the energy functional with isospin asymmetry is
particularly uncertain\cite{baoan}. 
Defining the total baryonic density and the isospin asymmetry as a function of the proton and neutron densities as $\rho=\rho_n+\rho_p$, $\delta=(\rho_n-\rho_p)/(\rho_n+\rho_p)$, the nuclear matter energy per nucleon in an asymmetric system can be expanded in powers of the asymmetry as
\begin{equation}
e(\rho,\delta)=e_0(\rho)+e_{sym}(\rho)\delta^2 +O(\delta^4)
\label{esym}
\end{equation}
where the symmetry energy is defined as the curvature of the energy functional in the asymmetry direction, $2e_{sym}=\partial^2e(\rho,\delta)/\partial \delta^2$.

The behavior of the symmetry energy in a number of modern phenomenologic or microscopic approaches reproducing a large set of mass and charge nuclear physics data for stable as well as exotic nuclei is shown in Figure 4\cite{fuchs2}. All of these approaches fulfill the incompressibility constraint for symmetric nuclear matter, with the NL3 parameter set of the relativistic mean-field approach being at the stiffest edge of the constraint ($K_\infty=272$ MeV).
\begin{figure}
\includegraphics[width=5.in,height=4.in,clip]{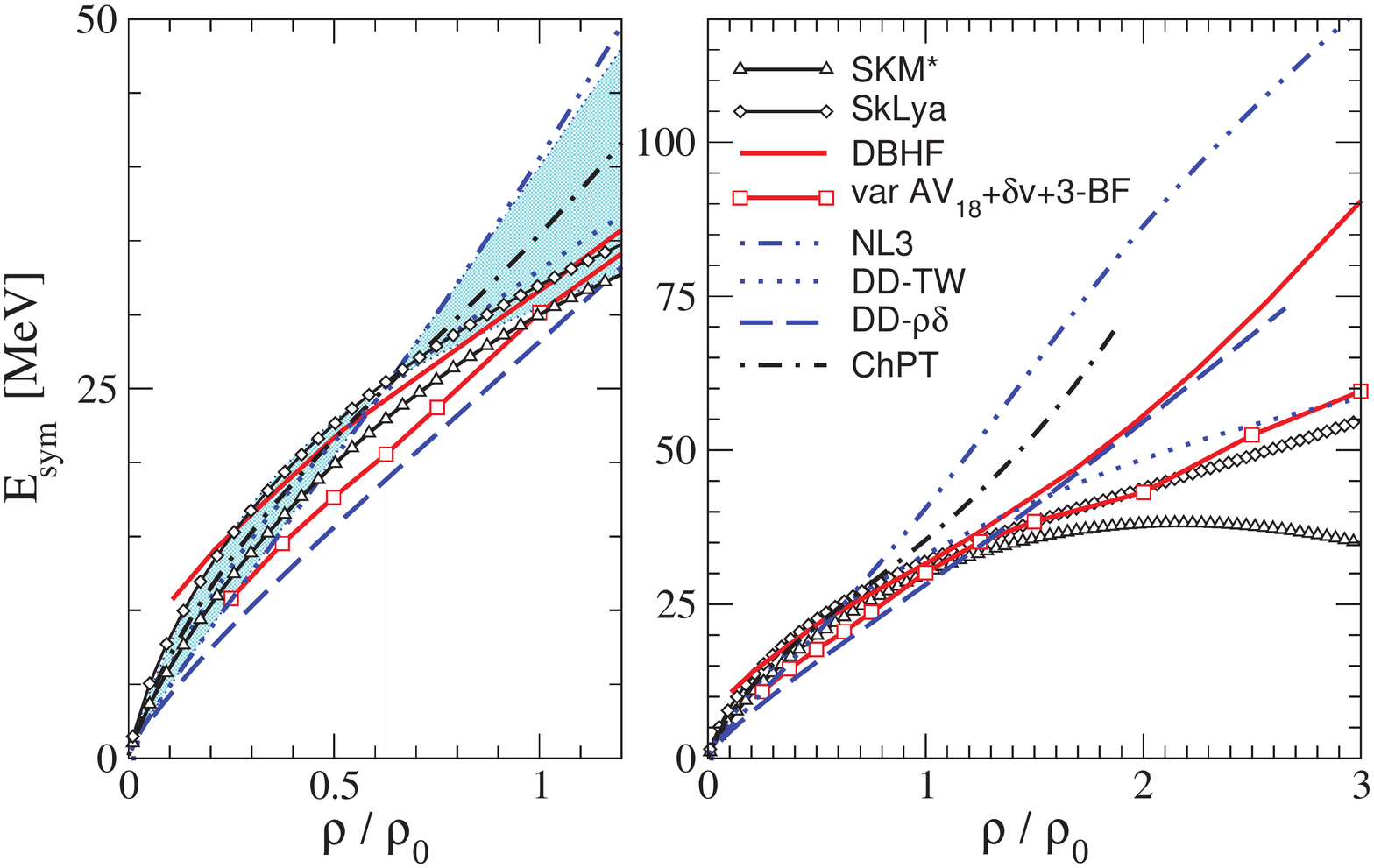}
\caption{ (Color online) Nuclear matter symmetry energy as a function of the baryonic density in different phenomenologic and microscopic approaches.  Figure taken from ref.\cite{fuchs2}.
 .}
\label{fig4}
\end{figure}
We can see from this picture that the density dependence behavior of the symmetry energy is poorely constrained even at subsaturation densities. 

This density dependence is however the most important equation of state ingredient in determining a number of phenomena in neutron star physics. This is easy to understand considering that, if higher order terms in the asymmetry can be neglected in eq.(\ref{esym}), the symmetry energy can be equivalently defined as
the difference between the energy per baryon of pure neutron matter and symmetric matter  at the same total density:
\begin{equation}
e_{sym}(\rho)\approx e(\rho,\delta=1)- e(\rho,\delta=0).
\label{esym2}
\end{equation}
This means that, for a very neutron rich system as a neutron star where typical proton fractions are believed to be in the range $x_p\approx 0.1-0.15$, supposing a constrained symmetric matter equation of state the whole residual uncertainties in the isovector channel come from the symmetry energy.
The importance of the symmetry energy in neutron star physics is developed at length in many recent reviews\cite{baoan}. Let us mention that the density dependence of the symmetry energy is a key quantity in determining the mass and density profile of neutron stars, as well as their equilibrium proton fraction. This information in turn influences the neutrino emission probability and determines the cooling rate of proto-neutron stars which can be measured from X-ray bursts.

Because of the uncertainties in extracting the isovector properties of the isovector part of the nuclear equation of state, in parallel with the theoretical progress, there has been  a strong effort during the last decade in trying to constrain the symmetry energy from different experimental observables, including mass fits with sophisticated mass formulas (FRDM)\cite{moller}, isospin diffusion (HIC)\cite{tsang,sun} and transverse flow\cite{kohley} in intermediate energy heavy-ion collisions, isobaric analog states (IAS)\cite{lee}, neutron thickness measurements via polarized proton elastic scattering(Pb(p,p))\cite{zenihiro},
pygmy dipole resonance (PDR)\cite{carbone}, and parity-violating electron-nucleus scattering\cite{roca}.

In order to better visualize the density dependence, it is useful to expand the symmetry energy in a Taylor series 
around the saturation density:
\begin{equation}
e_{sym}(\rho)=S_0+\frac{1}{3}L\frac{\rho-\rho_0}{\rho_0}+\frac{1}{9}K_{sym}\frac{
(\rho-\rho_0)^2}{{\rho_0}^2}+O\left [ \frac{(\rho-\rho_0)^3}{\rho_0^3}\right ].
\label{esym3}
\end{equation}
 Within a given set of models the different coefficients of the expansion are strongly correlated, such that the determination of two parameters is generally sufficient to determine the behavior of the functional, at least in the density interval probed by the constraint. 
\begin{figure}
\includegraphics[width=0.48\columnwidth,height=3.in,clip]{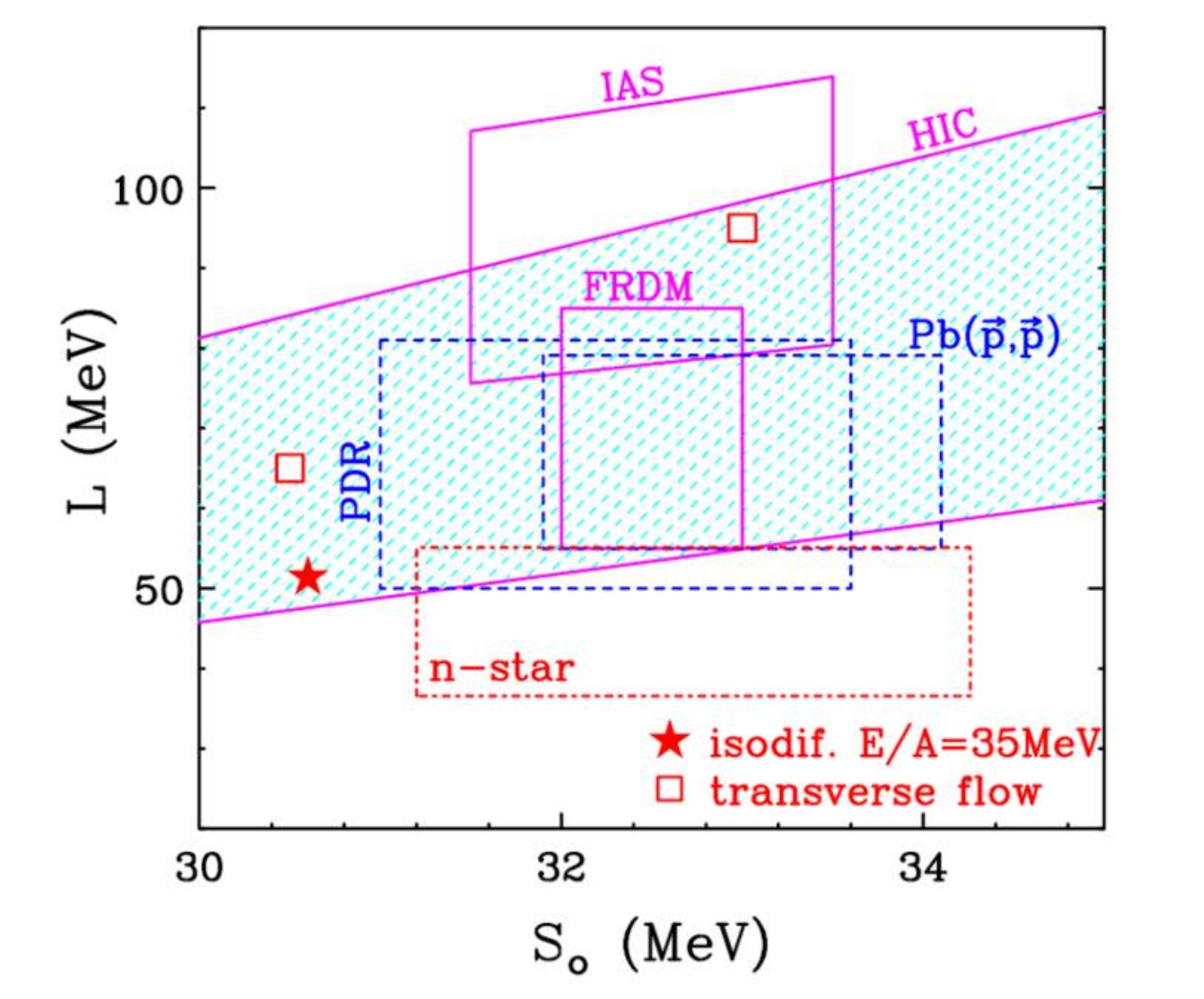}
\includegraphics[width=0.48\columnwidth,height=3.in,clip]{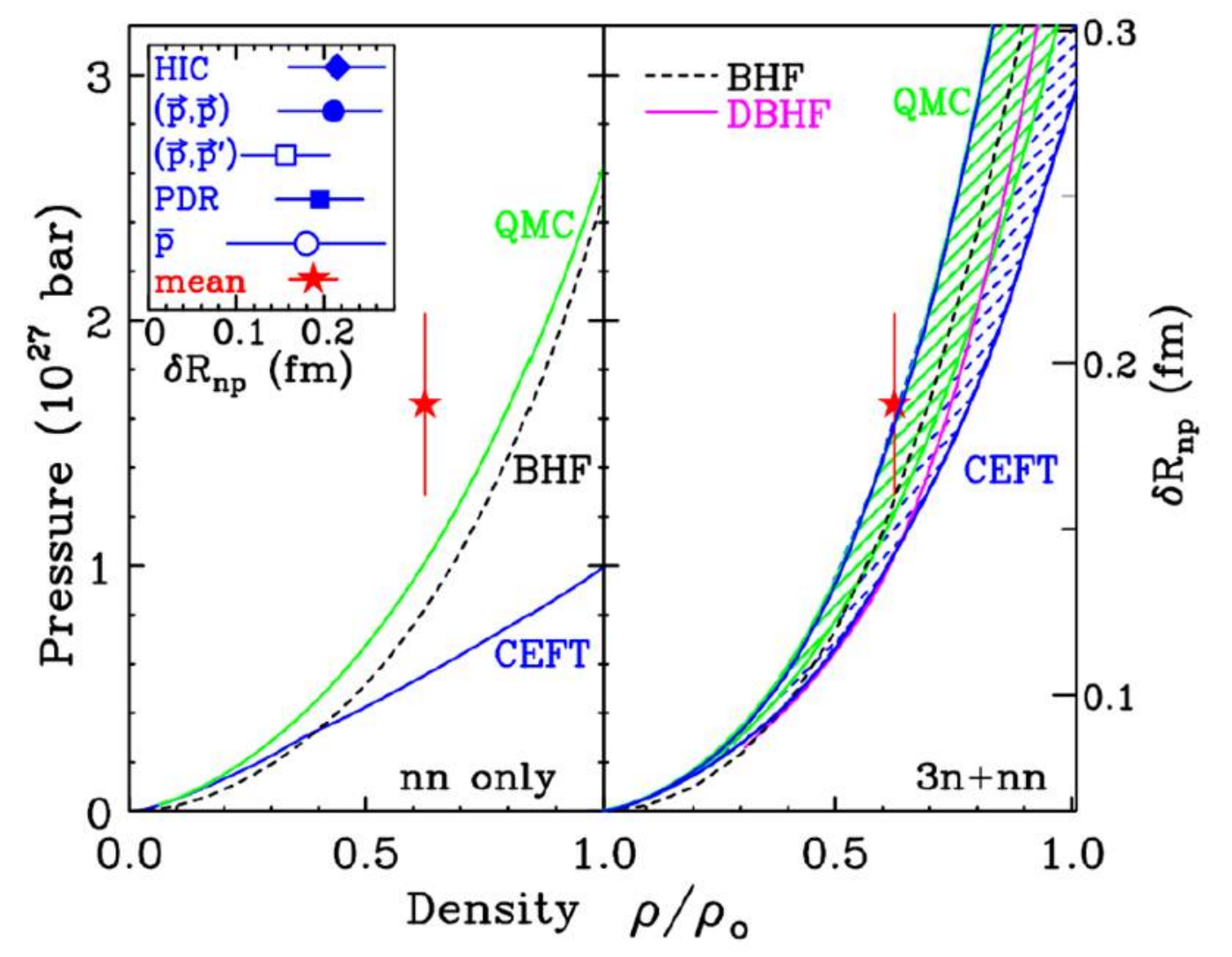}
\caption{ (Color online) Left part: present constraints on the symmetry energy and its slope at saturation with different experimental observables (see text). Right part: pressure as a function of the density in pure neutron matter with and without three-body interaction in Brueckner-Hartree-Fock (BHF), Chiraal effective field theory (CEFT) and quantum Monte Carlo (QMC) calculations. The associated neutron thickness in Lead is also given. The star represents the weighted average of different pressure, ore equivalently neutron thickness estimations in various experimental observables as given by the inner panel. Both figures are  taken from ref.\cite{nusym}.Copyright (2012) by the American Physical Society. 
 .}
\label{fig5}
\end{figure}
The latest constraints to the first two coefficients in the density expansion of the symmetry energy are shown in the left part of Figure \ref{fig5}. The convergence of the different results is impressive, considering that they come from completely independent observables and analyses. 

It is however important to stress that all of these results come from the confrontation of experimental data to to well controlled, but not ab-initio nuclear model. Because of that, the systematic error bars associated to the different estimations are difficul to evaluate and the recent results of Figure \ref{fig5} will have to be analyzed with care before a definite conclusion on the symmetry energy coefficients and their associated error bars can be safely taken.

The lower box in the left part of Figure \ref{fig5} labelled `n-star'\cite{steiner}  is a symmetry energy constraint that does not come from laboratory experiments, but is based on the most recent measurements of   light neutron stars where both mass and radius can be measured simultaneously through X-ray data. In order to convert such a measurement into a symmetry energy coefficient it is necessary to make some assumptions on the dynamics of X-ray bursts, and on the emissivity of the stellar surface. The uncertainty in these estimations is propagated to give an incertainty on the symmetry energy parameters, which is given by the error bars shown in Figure \ref{fig5}.

If this estimation is reliable, this implies that astrophysical observations are becoming competitive with laboratory experiments in order to get information on the nuclear energy functional. This conclusion should however be taken with caution, because nuclear physics hypotheses, together with star and atmosphere modelling are needed in order to produce the given estimation. In particular, in order to obtain the number given in Figure \ref{fig5},  it is assumed that  the equation of state of symmetric matter is completely determined by the incompressibility at saturation, that the equation of state of neutron matter is reliably given at by quantum Monte Carlo calculations with nucleons only\cite{gandolfi} up to density $\rho=0.4$ fm$^{-3}$ and can be continuated by polytropic expressions for higher densities, and that the parabolic approximation eq.(\ref{esym2}) is valid for all densities and asymmetries. All these assumptions need to be critically verified before a reliable error bar can be estimated.

Even within this word of caution, it is interesting to observe that this astrophysical constraint is the only one in Figure \ref{fig5} which uses information from densities much higher that the saturation density. The marginal compatibility of this estimation with respect to the laboratory experiments which all probe uniquely the sub-saturation density regime might suggest that new phenomena occur at supersaturation density.

If we limit ourselves to the low density region where the nuclear physics calculations are the most reliable,
the experimental estimation of the nuclear symmetry energy gives important information and constraints on the nuclear interaction, particularly the importance of three-body forces. This can be seen in the right part of Figure \ref{fig5}\cite{nusym} which shows the pressure of pure neutron matter as a function of the density in different microscopic calculations taking into account neutron-neutron interactions only, or alternatively including also three-body forces. Using the approximation eq.(\ref{esym2}), the symmetry energy can be converted into an estimation  of the neutron pressure. This is given by the star, which represents a weighted average of the different estimations from nuclear laboratory experiments. A systematic uncertainty on the density explored in these observables certainly exists, but is difficult to evaluate and is not represented in the figure. We can see that the measured symmetry energy is only compatible with calculations including three-body forces. If the present error bar is  not yet sufficiently small to determine the still highly uncertain three-neutron interaction, which makes the extrapolation towards supersaturation density somewhat hazardous, it is clear that it represents an important piece of evidence of the existence of such high order terms even at very low densities. It is important to stress that none of the microscopic models shown in the right part of Figure \ref{fig5} was used in the analysis of the symmetry energy data.

\section{Equation of state at finite temperature}
Since the early days of the equation of state modelization it was clearly recognized that, because of the strong similarities between the effective nucleon-nucleon potential and molecular interactions, the phase diagram of diluted symmetric nuclear matter should present a first order phase transition of the liquid-gas type, terminating at high temperature and density in a critical point\cite{siemens}. This conjecture has been confirmed in all phenomenological\cite{ducoin,rios} or microscopic\cite{typel} modelizations of nuclear matter with realistic effective interactions. 

\begin{figure}
\includegraphics[width=0.43\columnwidth,height=3.in,clip]{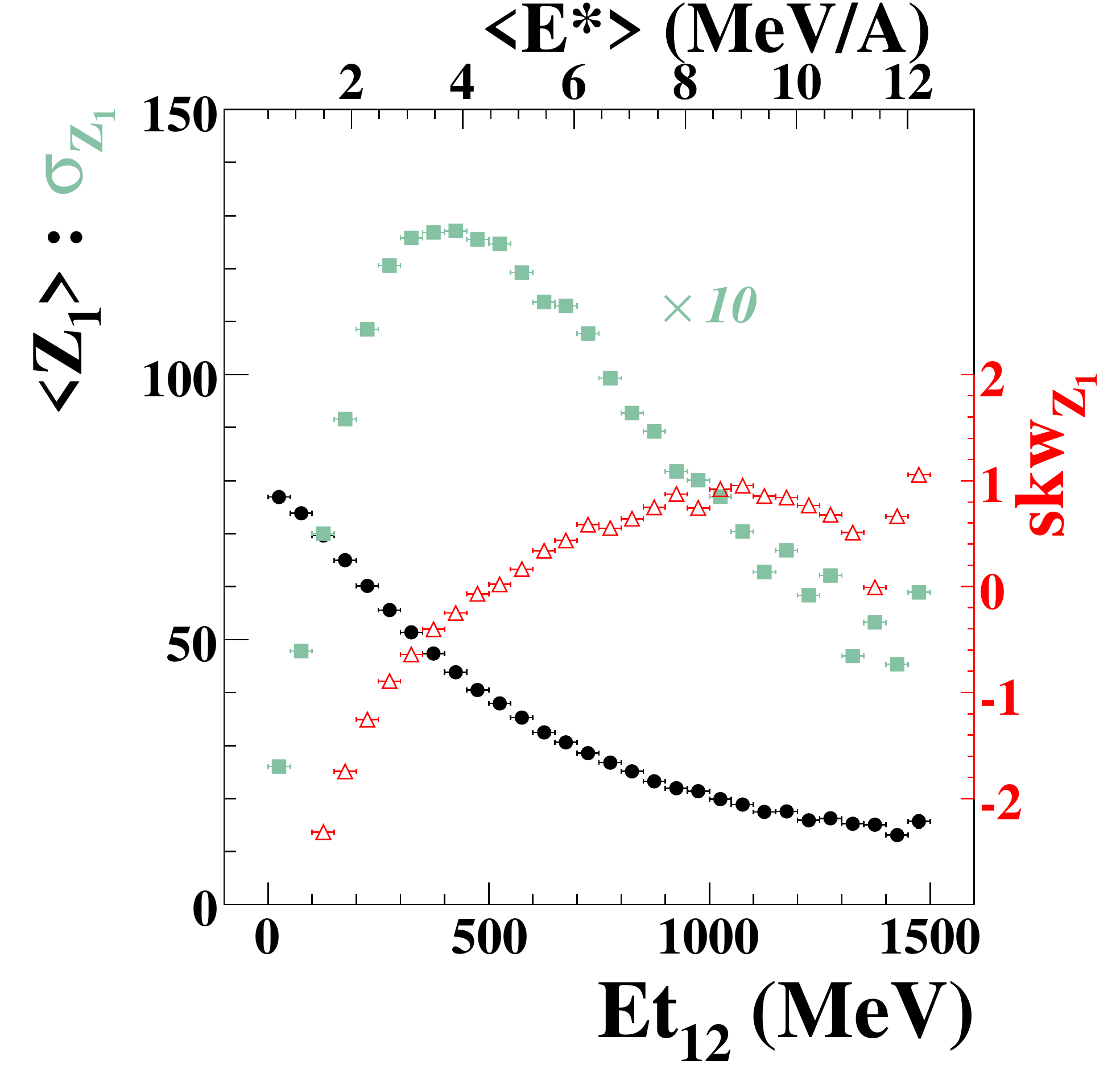}
\includegraphics[width=0.43\columnwidth,height=3.in,clip]{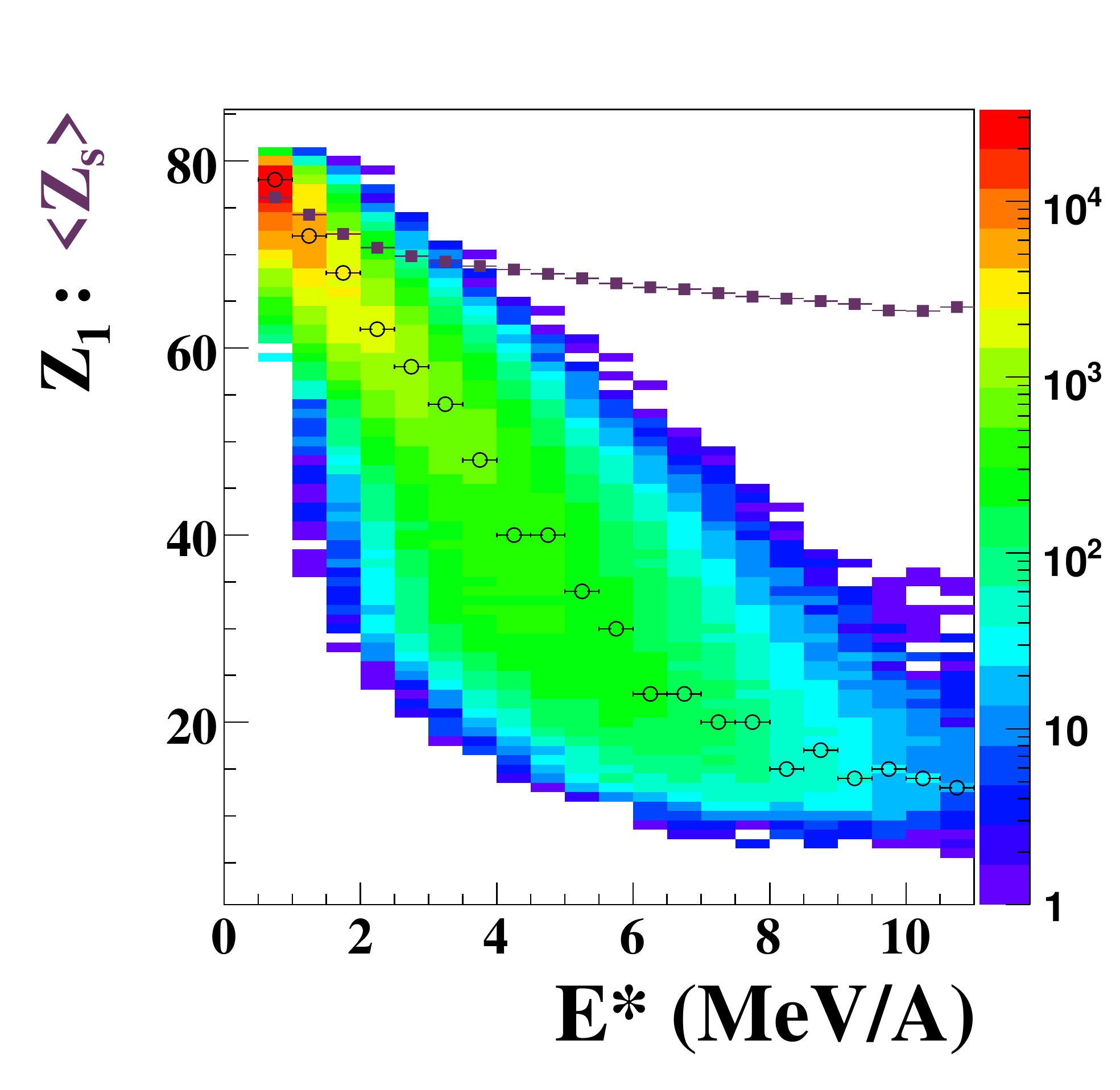}
\includegraphics[width=0.9\columnwidth,height=4.in,clip]{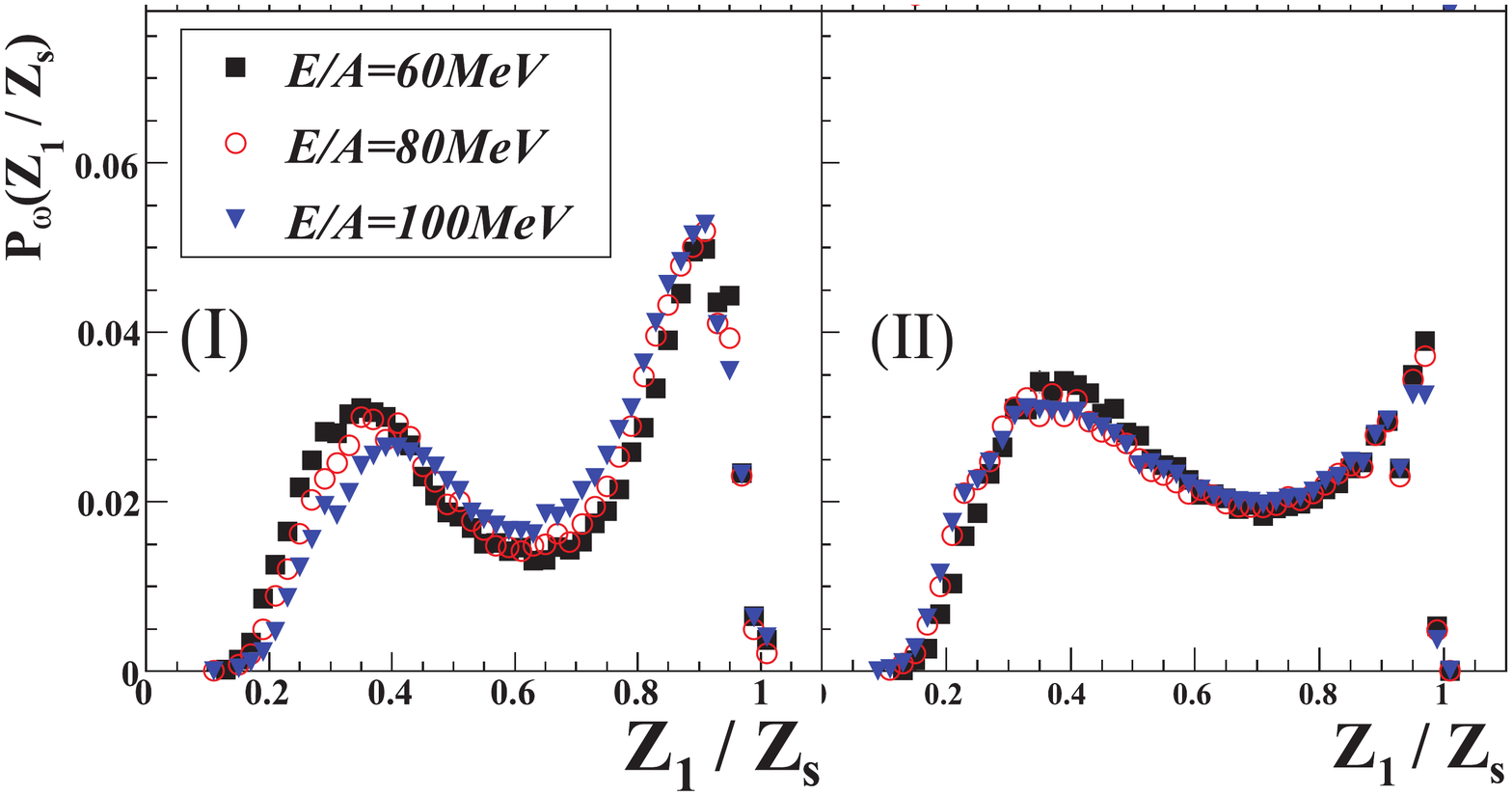}
\caption{ (Color online) Properties of the fragmentation of   gold quasi-projectiles produced in peripheral Au+Au
collisions with the INDRA apparatus. Upper left:average (dots), standard deviation
(squares) and skewness (triangles - right Y-axis) of the distribution
of the heaviest fragment  as a function of the light charged particles transverse energy at 80 MeV/nucleon . Upper right: correlation between the charge of the heaviest fragment and the calorimetric excitation energy.
The open squares indicate the most probable Z1 values. The average total source size Zs is given by the full symbols. Lower part: measured distribution of the charge of the largest fragment normalized to the charge of the source and to the number of collected events in each excitation energy bin
at three different bombarding energies.  
The left (right) side shows distributions obtained with two different
data selection methods.  Figures are  taken from ref.\cite{bimo}. Copyright (2009) by the American Physical Society. 
 }
\label{fig6}
\end{figure}

Experimental evidence and characterization of this phase transition has been very actively seeked for in the last decades. Many different signals have been proposed and measured\cite{finn,cneg,elliott,pseudo} during the years, but the  most compelling and model independent signature of a first order phase transition in nuclear multifragmentation is the recent observation of a bimodal pattern in the fragmentation of Au quasi-projectiles evidenced by the INDRA collaboration\cite{bimo}.  The corresponding experimental data are shown in Figure \ref{fig6}.

At the transition point of a first order phase transition, the distribution of the order parameter
in the corresponding finite system presents a characteristic bimodal behavior. The bimodality 
physically corresponds to the simultaneous presence of two
different classes of events, which, if the system was  at the thermodynamic limit, could be interpreted as 
 phase coexistence.
In the case of nuclear multi-fragmentation, the most natural observable to analyze as a potential
order parameter is the size of the heaviest cluster produced in each collision event. Indeed this
observable is known to provide an order parameter for a large class of transitions or critical phenomena
involving complex clusters, from percolation to gelation, from reversible to irreversible
aggregation. The bimodal pattern observed in Fig.\ref{fig6} is therefore a compelling evidence of a phase transition, and reaccelerated exotic beams at intermediate energies will be very important to settle how this observation evolves far from stability.

However the connexion of this fragmentation phase transition to the liquid-gas transition of ideal nuclear matter is very difficult to establish. Not only finite size effects dominate the statistical mechanics of such small systems, but also the role of the Coulomb interaction is certainly not negligible in the fragmentation process, and it cannot be disentangled from the efffect of the nuclear interactions. Moreover it is probable that thermal equilibrium is not achieved in these collisions, which makes the transition observed very loosely connected to the expected behavior in the bulk.

Because of these limitations, the present approach to the equation of state of nuclear matter at finite temperature is similar to the ground state strategy developed in the previous chapter. Namely, the phase diagram of bulk matter is constructed  within theoretical models  based on nuclear functionals which are constrained by appropriate experimental data.

At first sight the finite temperature properties of bulk asymmetric matter might look very disconnected to the experimentally accessible excited state properties of finite exotic nuclei, and one can doubt that valuable constraints can be obtained from such data. However, as we will develop in the next chapter, finite temperature nuclear matter as it is formed in supernova explosions and the cooling of proto-neutron stars is in reality essentially formed of excited finite neutron rich nuclei, meaning that the experimental information from exotic beams is directly relevant to the astrophysical problem. 

\section{The specificity of stellar matter }

As we have already pointed out, the theoretical idealization of nuclear matter consists in completely neglecting any possible Coulomb correlation. In the physical situation of stellar matter however, charge neutrality is achieved by the screening effect of electrons on the proton charge. Because of the very different mass between electrons and protons, the compressibility of electron and proton matter is very different, which induces Coulomb effects that drastically modify the liquid-gas phase transition associated to uncharged nuclear matter. Specifically, in all density and temperature conditions relevant for neutron star physics, the electron charge can be safely considered as uniformly distributed \cite{maruyama}. This means that any baryonic density fluctuation (which is correlated to a proton density fluctuation because of the symmetry energy) induces a fluctuation in the electric charge. A well known consequence of the resulting Coulomb correlations is that  the low density phase at zero temperature is not a gas, but a Wigner crystal of nuclei immersed in the homogeneous electron background\cite{lattimer}. It is clear that such Coulomb effects do not disappear with increasing density and temperature, and it is a-priori not at all evident that a phenomenology equivalent to the one calculated for uncharged nuclear matter might at all be observed. However, guided by the uncharged nuclear matter example, 
the standard treatments currently used in most supernovae codes describe the dilute stellar matter at finite temperature in the baryonic sector as a statistical equilibrium between protons, neutrons, alphas and a single
heavy nucleus\cite{LS91,shen}. The transition to homogeneous matter in the neutron star core is supposed to be first order in these modelizations and obtained through a Maxwell construction in the total density at fixed proton fraction.

From the nuclear physics side it is well recognized that, since stellar matter is subject to the contrasting couplings of the electromagnetic and the strong interaction acting on comparable length scales because of  the electron screening, this should give rise to the phenomenon of frustration\cite{horowitz}, well-known in condensed matter physics\cite{campa}. Because of this, a specific phase diagram, different from the one of nuclear matter and including inhomogeneous components, is expected in stellar matter\cite{noi}. 

\begin{figure}
\includegraphics[width=0.43\columnwidth,height=4.in,clip]{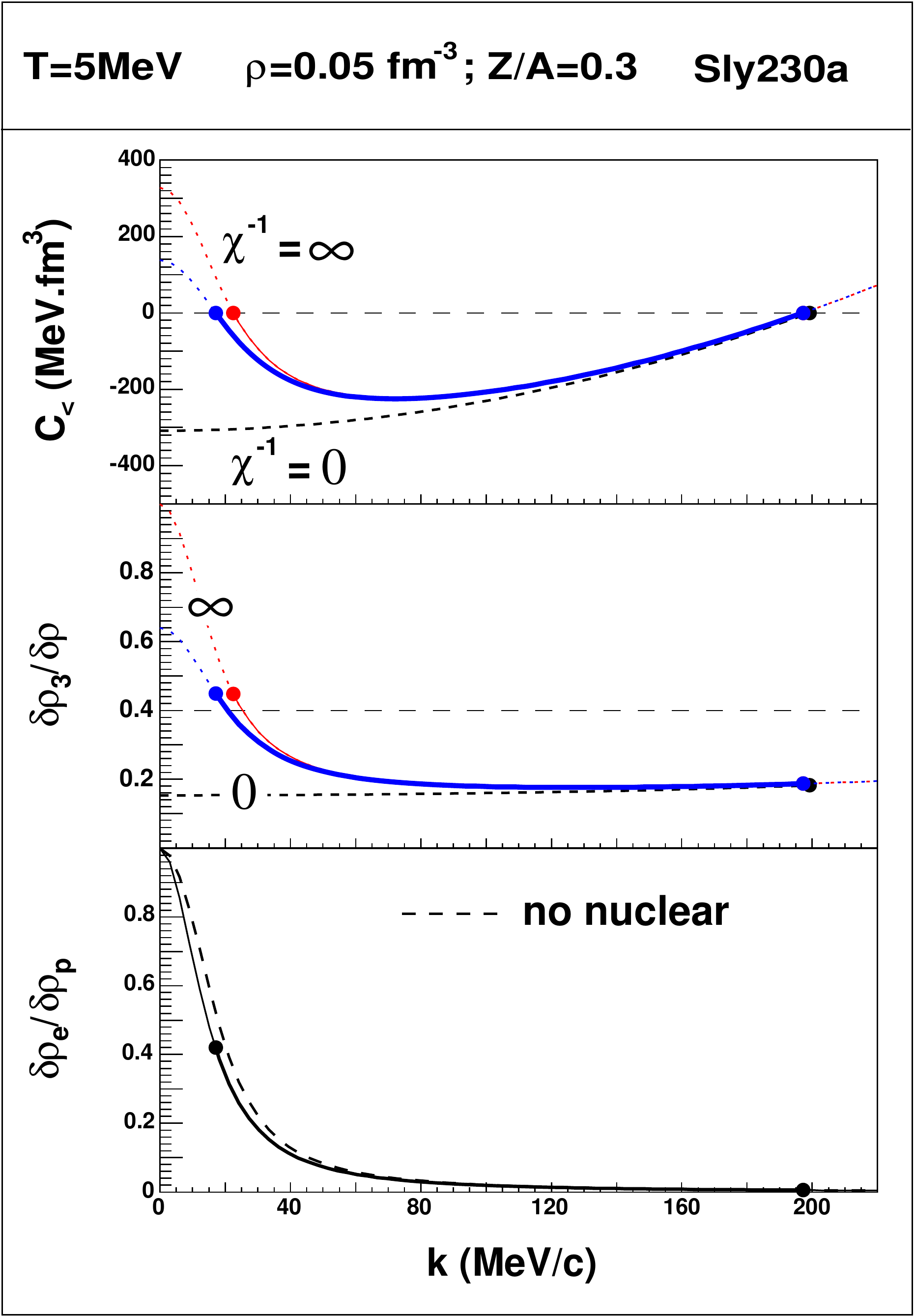}
\caption{  (Color online) Eigen-mode of the minimal free-energy curvature for $\rho$ = 0.05 fm−3, Z/A = 0.3, T = 5 MeV, as a function of the wave number k, calculated with the Sly230a Skyrme interaction. Top: eigen-value. Middle: associated eigen-vector in the nuclear-density plane. The curves are compared to two limiting cases corresponding to a zero and infinite incompressibility of the negatively charged gas. When the curvature is negative (full lines) this eigen-vector gives the phase-separation direction. The dots give the points of zero curvature. The dashed line gives the direction of constant Z/A. Bottom: same as the middle part, in the plane
of proton and electron density. The dashed line gives the eigenvector when the nuclear force is zero.  Figure  taken from ref.\cite{ducoin1}.
 }
\label{fig7}
\end{figure}

The quenching of the first order phase transition due to Coulombic effects is illustrated in Figure \ref{fig7} in a mean-field calculation with realistic effective interactions\cite{ducoin1}.
In order to determine the possible first order phase transition in stellar matter, the instability of such matter with respect to a density fluctuation can be studied computing the eigenvalues and eigenvectors of the free energy curvature matrix 
\begin{equation}
C_{ij}=\frac{\partial^2 f}{\partial\delta\rho_i\partial\delta\rho_j}
\end{equation}
once independent proton, neutron, and electron fluctuations $(q=n,p,e)$
\begin{equation}
\delta \rho_q = A_q e^{i\vec k \cdot \vec r} + c.c.,
\end{equation}
are imposed in a given thermodynamic condition defined by a density, temperature, and proton fraction\cite{pethick}. A negative eigenvalue associated to a given wavelength signals that a fluctuation characterized by the associated eigenvector will be spontaneously amplified, giving rise to cluster formation if the wavelenght is finite, or phase separation for infinite wavelengths.  This smallest eigenvalue $C_<$  is shown as a function of the wave number in Figure \ref{fig7}. We can see that the eigenvalue is positive at k=0, meaning that the phase transition is quenched and replaced by cluster formation in stellar matter.

This Coulomb frustration effect has been confirmed by different microscopic models\cite{watanabe} .
A continuous evolution from the Coulomb lattice to an homogeneous nuclear fluid, passing through the formation of clusters of different sizes and strongly deformed dishomogeneous structures close to the saturation density has been reported, both at zero and at finite temperature. 
These calculations are numerically very heavy and a complete thermodynamic characterization of stellar matter under the frustrated phase transition has not been done yet. Such a task can however be performed in the phenomenological  so-called nuclear statistical equilibrium (NSE) approaches,  which treat the bound states of
nucleons as new species of quasiparticles\cite{philips}.
Within NSE, the baryonic component of the stellar matter is regarded as a statistical equilibrium of neutrons and
protons, the electric charge of the latter being screened by a homogeneous electron background.
As a first approximation, one can consider that the system of interacting nucleons is equivalent to a system of noninteracting clusters, with nuclear interaction being completely exhausted by clusterization. This simple model can describe only diluted matter at $\rho\ll\rho_0$ as it can be found in the outer crust of neutron stars,
while nuclear interaction among nucleons and clusters has to be included for applications at higher density,when the average interparticle distance becomes comparable to the range of the
force. 
A simple possibility\cite{raduta} is to take into account  interactions among composite clusters  in the simplified form of a hard sphere excluded volume, and interactions among nucleons  in the self-consistent Hartree-Fock approximation with a phenomenological realistic energy functional.

\begin{figure}
\includegraphics[width=0.8\columnwidth,height=5.in,clip]{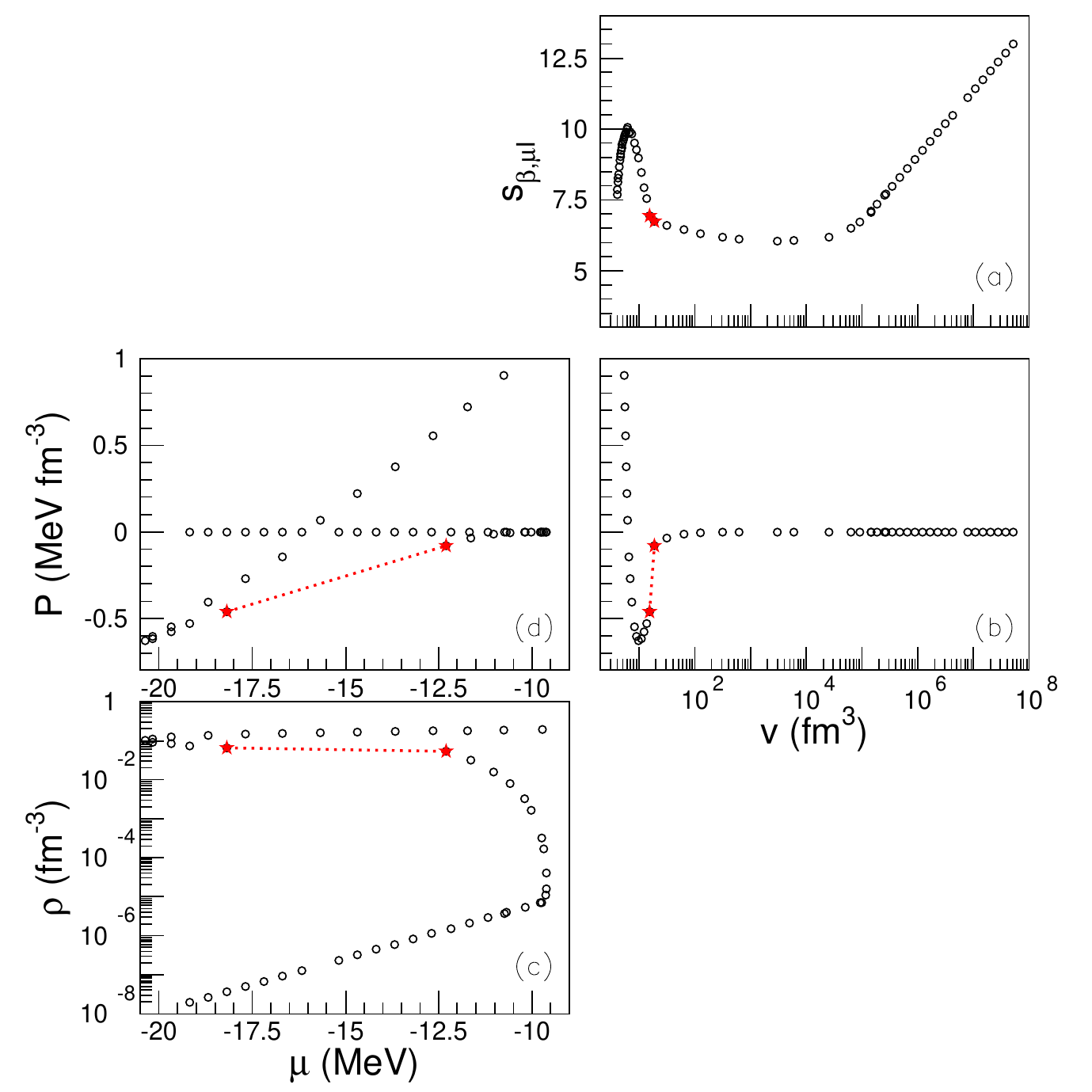}
\caption{ (Color online) Constrained entropy (a), pressure [(b) and (d)] and chemical potential [(c) and (d)] evaluated in the extended NSE model in the canonical ensemble at T = 1.6 MeV and asymmetry chemical potential $\mu_I$ = 1.68 MeV.  Figure  taken from ref.\cite{raduta2}. Copyright (2012) by the American Physical Society. 
 }
\label{fig8}
\end{figure}

The thermodynamics of the extended NSE model is presented in Figure \ref{fig8}. We can see that the equations of state do not present any plateau as it would have been the case for a first order phase transition.
More surprising, the entropy presents a convex intruder, the behavior of the equations of state is not monotonic and a clear backbending is observed, qualitatively similar to the phenomenon observed in  phase transitions in finite systems\cite{gross}.  It is interesting to remark that similar behaviors, with non-monotonic equations of state and discontinuities in the intensive variables, are systematically observed in phase transitions with long-range interactions\cite{campa}. A consequence of the backbending in the equation of state is that this unusual thermodynamics can only be observed in the canonical ensemble. Indeed if the baryon chemical potential was controlled as it is the case in standard NSE, in the region of the backbending the multiple evaluation of the chemical potential would lead to keep only the solution of minimal free energy. This means that the whole backbending region would be jumped over and one would observe a density discontinuity, that is a first order phase transition. This inequivalence of statistical ensemble is a characteristic feature of phase transitions with long range interactions. Different model applications have indeed shown fingerprints of ensemble inequivalence
\cite{campa}, but phenomenological applications are scarce. The NSE calculation  of Figure \ref{fig8} shows that the inhomogeneus baryonic matter which is produced in the explosion of core-collapse supernova and in neutron stars is an example of a physical system which displays this inequivalence.

\begin{figure}
\includegraphics[width=0.45\columnwidth,height=3.in,clip]{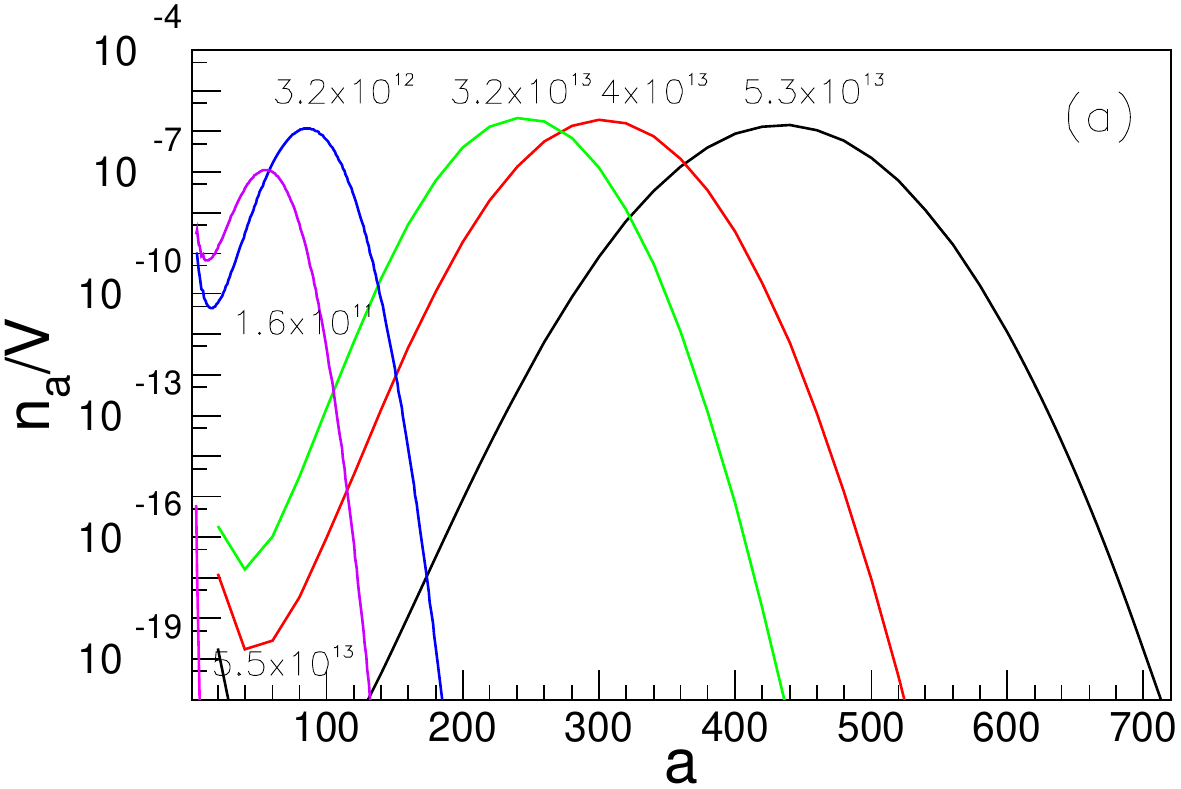}
\includegraphics[width=0.45\columnwidth,height=3.in,clip]{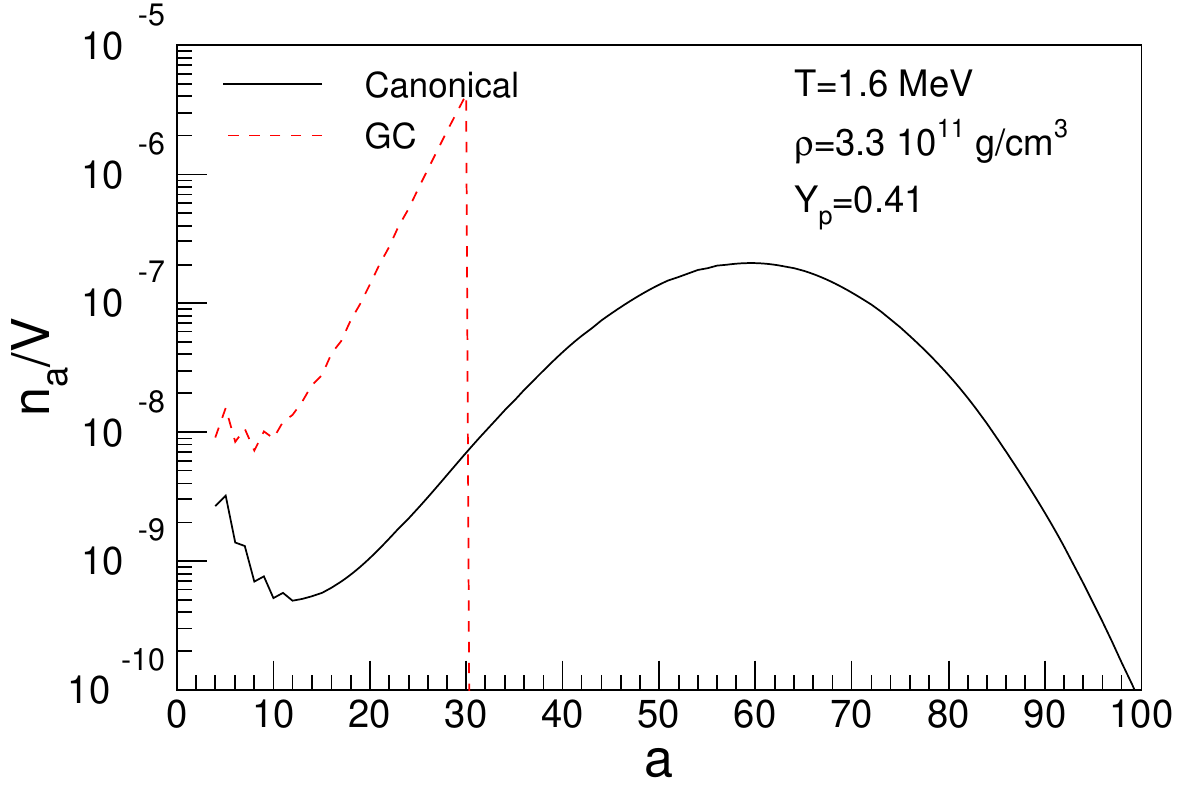}
\caption{  (Color online) Left side: Cluster distributions as a function of the density (expressed in $g/cm^3$) of the extended NSE model in the ensemble inequivalence region in the same thermodynamic conditions than in Figure \ref{fig8}. 
Right side: Comparison between canonical (full line) and grand-canonical (dashed line) predictions for the cluster distribution in a specific thermodynamic condition relevant for supernova dynamics. Figure  taken from ref.\cite{raduta2}. Copyright (2012) by the American Physical Society. 
 }
\label{fig9}
\end{figure}
From the physical viewpoint, the correct modelization is the canonical one. Indeed  if the grancanonical phase coexistence solution was the preferred response at equilibrium inside the inequivalence region, such solution would have been found in a canonical calculation where the coexistence density is imposed. On the contrary, once the density is fixed, a dishomogeneous clusterized solution is found in agreement with the expected phenomenology associated to the Coulomb frustration.
This is demonstrated in the left part of Figure \ref{fig9}, which shows the cluster distribution in the inequivalence region. As a function of the density, the average cluster size is continuously changing and can never be assimilated to a portion of the fluid phase. The right part of the same figure compares the grandcanonical and canonical cluster distribution in a thermodynamic situation relevant for supernova physics. We can see that accounting for the correct thermodynamics has important consequences on the matter composition. Since the cluster abundancies determine the electron capture rate, which in turn is a capital ingredient for size of the homologous core and the cooling process, it is clear that a control of the phase diagram is important for supernova physics. It is also interesting to remark that the most probable abundancies concern neutron rich nuclei which are accessible in laboratory experiments: a detailed knowledge of the mass, level density and electron capture cross section for these nuclei is thus needed to have a reliable description of the equation of state. 

\section{Conclusions}
In this contribution we have reviewed the main progresses of the past thirty years in the knowledge of the nuclear equation of state for moderate baryon densities and temperatures,  in the thermodynamic situation where matter is constituted by nucleonic degrees of freedom.

An intense combined experimental and theoretical effort has lead to quantitative as well as qualitative progress in the field. Thanks to the improvements of extended mean-field theories and transport equations, firm constraints have been put on the isoscalar part of the energy functional. Benchmark ab-initio calculations with quantum Monte Carlo methods constitute powerful constraints to the effective theories. New nuclear interactions derived from chiral perturbation theory start to establish a link with the underlying structure of QCD.
On the isovector side, the availability of new data on nuclear masses, transport and collective modes allows to fix the first constraints on the density dependence of the symmetry energy, pointing to an important role of three body interactions even at subcritical densities. The major challenge for future investigations concerns the high density part, for which exotic beams at relativistic energies will be needed.

From the conceptual viewpoint the progress is even more impressive.
Lead from the old liquid drop representation of nuclear structure, nuclear matter was viewed in the eighties as the bulk limit of the atomic nucleus, that is an idealized infinite extension of the internal part of heavy nuclei. In this sense complex quantum effects as shell structure and pairing were viewed as a finite size nuisance in the equation of state quest. Nowadays it is  admitted in the community that such homogeneous neutral medium does not exist, and that matter as it can be found in the crust of neutron stars and in the dynamics of exploding core-collapse supernovae is rather a collection of finite atomic nuclei that can be synthetized in the laboratory. This viewpoint modification leads to a great synergy between the nuclear physics and the astrophysical community which is expected to further develop in the next years.

\end{document}